\documentclass[11pt,twoside]{article}

\usepackage{asp2006}
\usepackage{epsf}
\usepackage{psfig}
\usepackage{lscape}

\begin{document}
\title{Primordial Mass Segregation in Starburst Stellar Clusters}   
\author{Sami Dib\altaffilmark{1,2,3}, Mohsen Shadmehri\altaffilmark{4,5}, Maheswar Gopinathan\altaffilmark{1,6}, Jongsoo Kim\altaffilmark{1}, and Thomas Henning\altaffilmark{7}}   
\altaffiltext{1}{Korea Astronomy and Space Science Institute, 61-1, Hwaam-dong, Yuseong-gu, Daejeon 305-348, Korea; dib@kasi.re.kr; maheswar@kasi.re.kr; jskim@kasi.re.kr}
\altaffiltext{2}{on leave to: Service d'Astrophysique, CEA/DSM/DAPNIA/SAp, C. E. Saclay, 91191, Gif-sur-Yvette, Cedex, France}
\altaffiltext{3}{Lebanese University, Faculty of Sciences, Department of Physics, El Hadath, Beirut, Lebanon}
\altaffiltext{4}{School of Mathematical Sciences, Dublin City University, Glasnevin, Dublin 9, Ireland; mohsen.shadmehri@dcu.ie}
\altaffiltext{5}{Department of Physics, School of Science, Ferdowsi University, Mashhad, Iran}
\altaffiltext{6}{Aryabhatta Research Institute of Observational Sciences, Manora Peak, Nainital 263129, India}
\altaffiltext{7}{Max-Planck-Institut for Astronomy, K\"{o}nigstuhl 17, D-69117 Heidelberg, Germany; henning@mpia.de}
    
\begin{abstract} We present a model to explain the mass segregation and shallow mass functions observed in the central parts of dense and young starburst stellar clusters. The model assumes that the initial pre-stellar cores mass function resulting from the turbulent fragmentation of the proto-cluster cloud is significantly altered by the cores coalescence before they collapse to form stars. With appropriate, yet realistic parameters, this model based on the competition between cores coalescence and collapse reproduces the mass spectra of the well studied Arches cluster. Namely, the slopes at the intermediate  and high mass ends are reproduced, as well as the peculiar bump observed at $6$ M$_{\odot}$. This coalescence-collapse process occurs on short timescale of the order of one fourth the free fall time of the proto-cluster cloud (i.e., a few $10^{4}$ years), suggesting that mass segregation in Arches and similar clusters is primordial. The best fitting model implies the total mass of the Arches cluster is $1.45 \times 10^{5}$ M$_{\odot}$, which is slightly higher than the often quoted, but completeness affected, observational value of a few $10^{4}$ M$_{\odot}$. The derived star formation efficiency is $\sim 30$ percent which implies that the Arches cluster is likely to be gravitationally bound. 
\end{abstract}

\section{Introduction}   

Mass segregation is observed in many young stellar clusters, both compact and sparse, with the most massive stars being preferentially located in their central parts (e.g., Pandey et al. 1992; Malumuth \& Heap 1994;  Brandl et al. 1996; Hillenbrand \& Hartmann 1998; Fisher et al. 1998; Figer et al. 1999b, Sagar et al. 2001; Stolte 2002; Le Duigou \& Kn\"{o}delseder 2002; Sirianni et al. 2002; Gouliermis et al. 2004; Sharma et al. 2007). Among young clusters, the class of very dense clusters, harboring large numbers of massive stars and known as starburst clusters such as the Arches cluster, NGC 3603, the Quintuplet cluster, Westerlund 1 and 2 and R 136 offers a very challenging testbed for the theories of massive star formation not only because of the extreme mass segregation observed in their central parts, but also due to the fact that their radially integrated stellar mass function is observed to be top heavy (Moffat et al. 1994; Figer et al. 1999; Stolte et al. 2002; Stolte et al. 2005; Stolte et al. 2006; Kim et al. 2006; Harayama et al. 2007; Brandner et al. this volume). Similar trends are observed for the super star cluster NGC 1705-1 (Sternberg 1998). 

In particular, the structure of the young (age $\sim 1-2$ Myrs), and dense ($\sim 10^{5}$ stars/pc$^{3}$) Arches cluster which is located at a projected distance of only 25 pc from the Galactic center has been studied extensively over the past few years. Its mass function has been determined using a variety of space borne and ground based instruments in the near infrared such as the {\it Hubble Space Telescope (HST)} NICMOS camera (Figer et al. 1999a), the Hokupa AO system on Gemini (Stolte et al. 2002), the NAOS/CONICA camera on the VLT (Stolte et al. 2005), and the NIRC2 instrument on Keck (Kim et al. 2006). Star counting in the inner annulus of the cluster (i.e., between the center and 1 time the core radius which is $\sim 0.2$ pc) is affected by incompleteness effects due to severe crowding, and the outer parts might still be contaminated by field stars. Thus, an interesting and well resolved area to test theoretical models is the second annulus. The converging results of these observations yield a very flat mass function with $\alpha \sim 1$ ($dN/d M=M^{-\alpha}$; The Salpeter value is $\alpha=2.35$, Salpeter 1955) in the central annulus of the cluster, a value of $\alpha \sim 1.7-1.9$ in the second annulus (i.e., between 1 and 2 times the core radius), and a nearly Salpeter-like or slightly steeper exponent in the outer areas.    
 
The mass segregation and shallow IMFs observed in starburst clusters such as Arches have been interpreted as the result of dynamical processes (Kim et al. 2006, Portegies-Zwart et al. 2007). Although dynamical processes will inevitably play a role in enhancing the fraction of massive stars in the central regions of a cluster, two points remain unclear: The first one is that the primordial IMF might be shallow in nature which leaves little room for the subsequent effects induced by dynamical friction. The second issue is that for young cluster as the Arches, it is not clear whether dynamical processes have time to play a significant role (see Freitag et al. this volume). Moreover, models based on mass segregation by dynamical processes do not reproduce, to date, some features of the Arches cluster mass function such as the peculiar bump at $\sim 6$ M$_{\odot}$.  In this work, we discuss a model to explain the observed mass function of the Arches and similar starburst clusters based on the coalescence of pre-stellar cores (PSCs) in the proto-cluster cloud and their subsequent collapse into stars. 

\section{The Coalescence-Collapse model}

\begin{figure} [!ht] 
\plotone{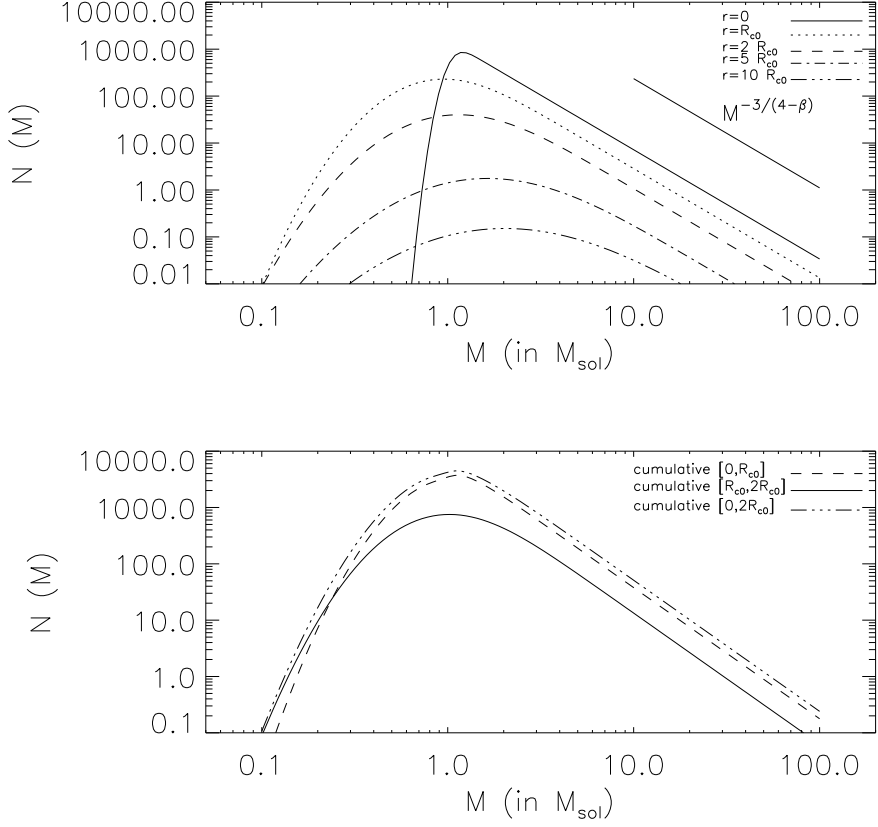}
\caption{Mass spectrum of Jeans unstable pre-stellar cores in shells of width=0.025 pc located at different distances from the cloud center (at 0, 1, 2,, 5, and 10 $R_{c}$; $R_{c}=0.2$ pc is the proto-cluster cloud core radius), where $\beta$ (here $\beta=1.8$) is the turbulent velocity power spectrum. Bottom: cumulative number of cores in the regions between [$0-R_{c}$], [$0,2~R_{c}$], and [$R_{c}-2~R_{c}$].}
\label{fig1}
\end{figure}

The coalescence model is discussed in detail in Dib et al. (2007a). Here, its basic features are summarized. PSCs are embedded in an isothermal molecular cloud (MC) (temperature is $T=10$ K), at different locations $r$ from the cloud's center. Both the PSCs and the MC are axisymmetric (PSCs are initially spherical but are likely to quickly flatten as time evolves). The MC and the PSCs have density profiles which are flat in their inner parts and follow $r^{-2}$ and $r^{-4}$ laws in their outer parts, respectively. However, due to their different locations in the MCs, PSCs of a given mass $M$ will have smaller initial radii $R_{p} (r,M)$ in the inner parts of the cloud. As time advances, the radius of the PSC will decrease due to gravitational contraction. The PSC contracts on a timescale, $t_{cont,p}$, which is equal to a few times its free fall timescale, and can be parametrized as:

\begin{equation} 
t_{cont,p}(r,M)= \nu ~ t_{ff,p}(r,M)= \nu \left( \frac {3 \pi} {32~G \bar{\rho_{p}} (r,M)} \right)^{1/2},
\label{eq1}
\end{equation}

\noindent where $\nu \ge 1$ and $\bar{\rho_{p}}$ is the radially averaged density of the PSC of mass $M_{p}$, located at position $r$ in the cloud. Thus, the time evolution of the radius of the PSC can be described by the following equation: 

\begin{equation} 
R_{p}(r,M,t)=R_p(r,M)~e^{-(t/t_{cont,p})}.
\label{eq2}
\end{equation}

Once the instantaneous radius of a PSC of mass $M_{p}$, located at position $r$ from the cloud's center is defined, it becomes possible to calculate its cross section for collision with PSCs of different masses. The cross section for the collision of a PSC of mass $M_{i}$ and radius $R_{i}$ with another of mass $M_{j}$ and radius $R_{j}$ and which accounts for the effect of gravitational focusing is given by: 

\begin{eqnarray} 
\sigma(M_{i},M_{j},r,t) = \pi \left[R_{p,i}(r,M_{i},t)+R_{p,j}(r,M_{j},t)\right]^{2} \nonumber \\
\times \left[ 1+\frac{2G (M_{i}+M_{j})} {2 v^{2} (R_{p,i}(r,M_{i},t)+R_{p,j}(r,M_{j},t))} \right].
\label{eq3}
\end{eqnarray}

As molecular clouds are unlikely to be virialized (e.g., Dib et al. 2007b), it is assumed that the cores relative velocity is not constant at a given position in the cloud and is rather better described by a Larson type relation $v(r)=v_{0} r(pc)^{\alpha}$ (Larson 1981; $v_{0}=1.1$ km s$^{-1}$), with a lower limit being the local thermal sound speed, which is uniform across the isothermal MC. As initial conditions for the PSCs mass distributions at different cloud radii, we adopt distributions that are the result of the gravo-turbulent fragmentation of the cloud, following the formulation given in Padoan \& Nordlund (2002). In these models, the mass distribution of cores is given by the following function:

\begin{equation}
N (r,M)~d~log~M = f_{0}(r)~M^{-3/(4-\beta)} \left[\int^{m}_{0} P(M_{J}) dM_{J}\right]d~log~M,
\label{eq4}
\end{equation}   

\noindent where

\begin{equation}
P(M_{J})~dM_{J}=\frac{2~M_{J0}^{2}}{\sqrt{2 \pi \sigma^{2}_{d}}} M^{-3}_{J} exp \left[-\frac{1}{2} \left(\frac{ln~M_{J}-A}{\sigma_{d}} \right)^{2} \right] dM_{J}.
\label{eq5}
\end{equation}

In Eq.~\ref{eq4}, $f_{0}(r)$ is the local normalization factor, and $\beta$ is the exponent of the turbulent velocity power spectrum and is related to $\alpha$ by $\beta=2 \alpha+1$. The dependence on the local dynamical and thermodynamical conditions is hidden in the terms $A$ and $\sigma_{d}$ which are given by $A=ln M_{J0}-\bar{ln~n'}$ and $\sigma_{d}=ln(1+{\cal{M}}^{2} \gamma^{2})$, where $M_{J0}$ is the Jeans mass at the mean density $n_{0}$, $n'$ the number density in units of $n_{0}$, $\cal{M}$ the local Mach number, and $\gamma$ a numerical factor found through numerical simulations to be $~\gamma=0.5$.

\section{Results}

With the initial conditions described in \S.~2, we follow the time evolution of the PSCs mass spectrum by solving the following equation of $N(r,M,t)$:

\begin{eqnarray} 
\frac{dN(r,M,t)}{dt}=0.5\times \eta(r) \times \nonumber \\
 \int^{\Delta M}_{M_{min}}~N(r,m,t)~N(r,M-m,t)~\sigma(m,M-m,r,t)~v(r)~dm \nonumber \\ 
 -\eta (r) N(r,M,t) \int^{M_{max}}_{M_{min}} N(r,m,t) \sigma(m,M,r,t) v(r)~dm,                            
\label{eq6}
\end{eqnarray}

\noindent where the first and second terms in the right hand side of Eq.~\ref{eq6} correspond to the rate of creation and destruction of a PSC of mass $M$, at location $r$, respectively (Nakano 1966; Shadmehri 2004). In Eq.~\ref{eq6}, $\Delta M=M-M_{min}$, and $\eta (r)$ is a coefficient which represents the coalescence efficiency, with $\eta \leq 1 $. This efficiency can be the result of various physical processes which can affect the coalescence of PSCs. For simplicity, we shall assume that $\eta$ is independent of position. In order to evaluate the transition from PSCs to stars, we compare, at each timestep, the local coalescence timescale to the local contraction timescale for PSCs of a given mass. The local coalescence timescale is $t_{coal}(r,M)=1/w_{coal}(r,M)$ where $w_{coal}$ is the coalescence rate (Elmegreen \& Shadmehri 2003): 

\begin{equation}
w_{coal}(r,M)=\frac{2^{1/2} v(r)}{V_{shell}(r)} \sum_{j=1}^{mbin} (R_{i}+R_{j})^{2} \left[1+\frac{2 G (M_{i}+M_{j})}{2 v^{2} (R_{i}+R_{j})} \right],
\label{eq7}
\end{equation}   

\noindent where $mbin$ is the number of mass bins, and $V_{shell}$ is the volume of the shell of width $dr$ located at distance $r$ from the MC's center. The contraction timescale is given by Eq.~\ref{eq1}. Whenever the local contraction timescale is shorter than the local coalescence timescale, PSCs are collapsed into stars. When a PSC collapses to form a star, a fraction of its mass is re-injected into the proto-cluster cloud in the form of an outflow. This mass loss is accounted for in a purely phenomenological way by assuming that the mass of a star which is formed out of a PSC of mass $M_{p}$ is given by M$_{\star}$=$\psi$ M$_{p}$, where $\psi \le 1$. Matzner \& McKee (2000) showed that $\psi$ can vary between $0.25-0.7$ for stars in the mass range $0.5-2$ $M_{\odot}$. As there is no evidence so far, for or against, whether this result holds at higher masses, it is assumed here that a similar fraction of the mass of a PSC is lost in the outflow independent of its mass. 

\begin{figure} [!ht]
\centering
\plotone{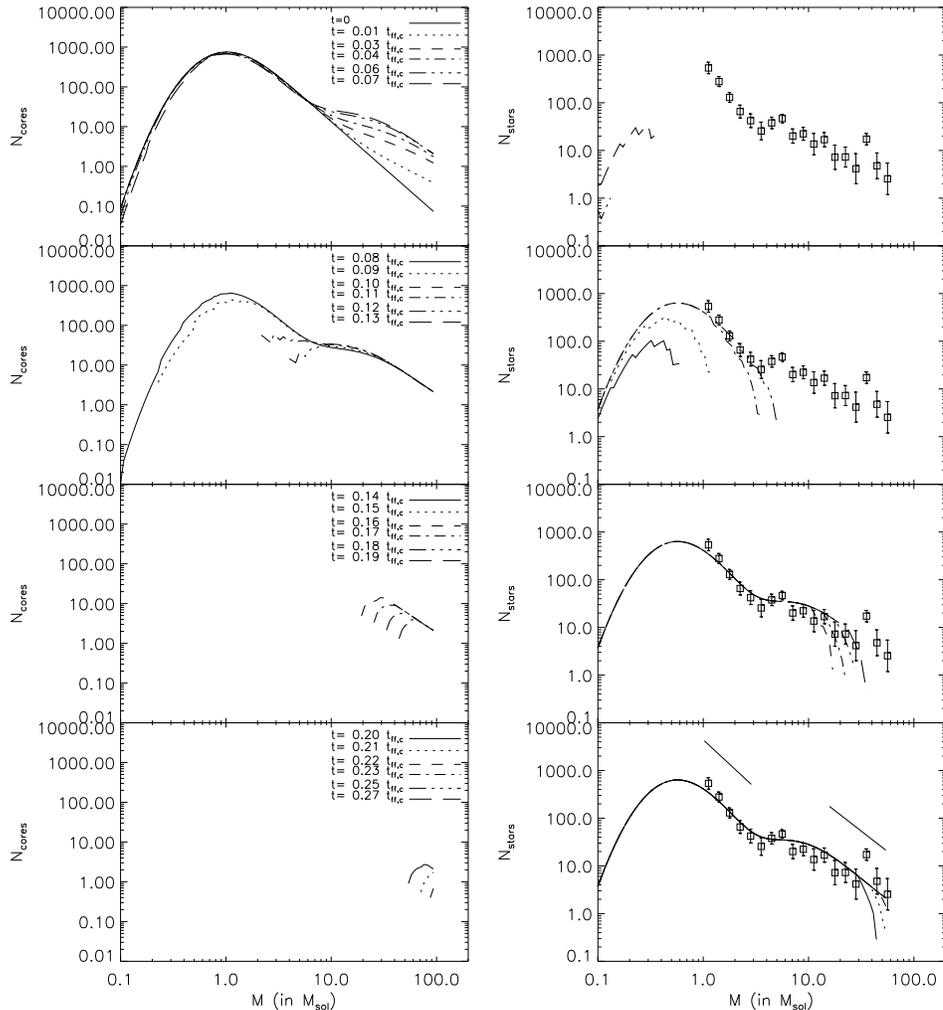} 
\caption{Time evolution of the pre-stellar cores mass function (left) and stellar mass function (right) in the region of the proto-cluster cloud between [$R_{c}-2~R_{c}$]. The stellar mass function is compared to that of the Arches stellar cluster mass function (Kim. et al. 2006; open squares). Fits to the simulated IMF (bottom right figure) yield slopes of -2.04$\pm~0.02$ and -1.72$\pm~0.01$ in the mass range of [1-3] M$_{\odot}$ and $\ge 15$ M$_{\odot}$, respectively, in very good agreement with the observations. Fits over-plotted to the data are shifted by one dex for the sake of clarity.}
\label{fig2}
\end{figure}

\begin{figure} [!ht]
\plotone{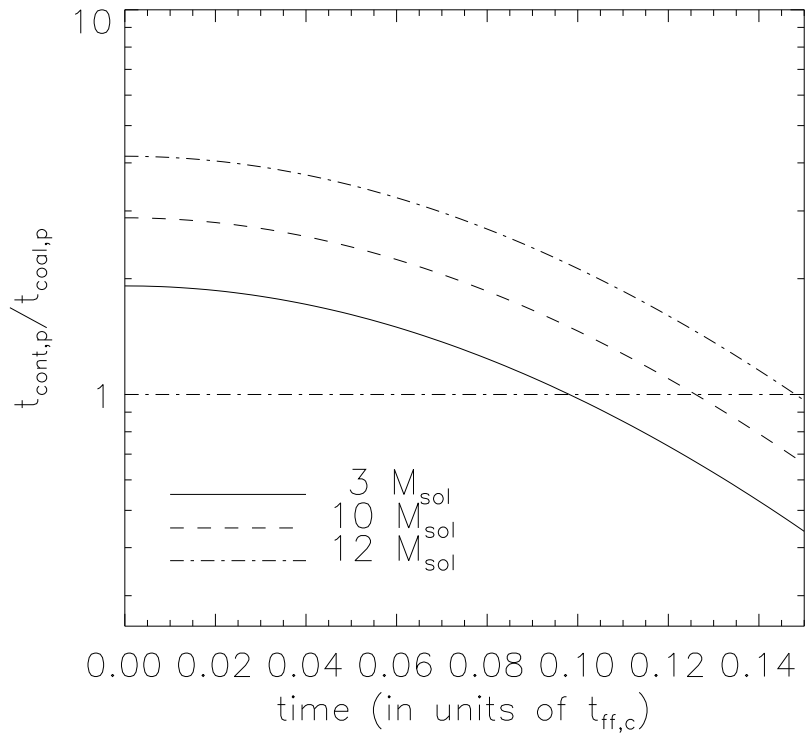} 
\caption{Time evolution of the ratio of the pre-stellar cores contraction timescale (equal to few times the free fall time) to the coalescence timescale for cores of masses, 3, 10 and 12 M$_{\odot}$ located at the center of the $R_{c}-2~R_{c}$ annulus in the proto-cluster cloud. Time is in units of the proto-cluster cloud free-fall time $t_{ff,c} \sim 3 \times 10^{4}$ years.}
\label{fig3}
\end{figure}

 Fig.~\ref{fig2} displays the time evolution of the cumulative PSC populations in the region [$R_{c0}-2~R_{c0}$]=[0.2-0.4] pc, which corresponds to the second annulus of the Arches cluster for a model with $\eta=0.5$, $\nu=10$, mass of the cloud $M_{c}=5 \times 10^{5}$ $M_{\odot}$, the radius and core radius of the cloud $R_{c}=5$ and $R_{c0}=0.2$ pc, respectively, an initial peak density of the PSCs $\rho_{p0}=10^{7}$ cm$^{-3}$, $\alpha=0.37$,$\psi=0.58$, and the fraction of the total cloud mass present in the cores $\epsilon=0.5$. In the initial stages, the most massive PSCs, that have a larger cross section, coalesce faster than the less massive ones, essentially by capturing the numerous intermediate mass PSCs and causing a rapid flattening of the spectrum at the high mass end. By $t \sim 0.07~t_{ff,c}$ ($t_{ff,c}=(3 \pi/32 G \bar{\rho_{c}})^{1/2} \sim 3 \times 10^{4}$ yr is the MC free fall timescale), a first generation of the smallest PSCs collapses to form stars. As time advances, more massive stars are formed in the shell (massive cores collapse later because of their lower average density) and in parallel the PSCs population decreases. By $t \sim 0.1~t_{ff,c}$ the intermediate mass PSCs which constitutes the largest mass reservoir for coalescence collapses into stars (see Fig.~\ref{fig3}). At this time, the turnover in the PSCs mass spectrum is located at $\sim 8-10$ $M_{\odot}$. Since the reservoir of intermediate mass objects is depleted, the remaining massive PSCs coalesce at a slower pace before they collapse. By $t \sim 0.25~t_{ff,c}$, all PSCs of different masses in the shell have collapsed and formed stars. Because of mass loss, the stellar IMF is shifted to lower masses (bump shifted to $\sim 6$ $M_{\odot}$). In summary, the resulting IMF is not very different from the PSCs mass spectrum after the initial and rapid stage of strong coalescence until $t \sim 0.01~t_{ff,c}$. In Fig.~\ref{fig2}, over-plotted to the numerical result is the cumulative mass spectrum of the Arches cluster in the annulus of [0.2-0.4] pc (Kim et al. 2006). The coalescence-collapse model agrees better with the observations than models based on mass segregation by dynamical friction. In particular, the bump at $\sim 6$ $M_{\odot}$ is reproduced. Fits to the stellar spectrum yield slopes of $\alpha= -2.04 \pm 0.02$ and $-1.72 \pm 0.01$ in the mass ranges of [$1-3$] $M_{\odot}$ and $\ge 15$ $M_{\odot}$, respectively, in very good agreement with observational values. With this set of parameters, the mass of the Arches clusters is found to be $1.45 \times 10^{5}$ which is slightly higher than the often quoted, completeness affect, observational value of a few $10^{4}$ M$_{\odot}$. This implies a star formation efficiency of (Mass of the cluster/Mass of the proto-cluster cloud)=($1.45 \times 10^{5}$ M$_{\odot}$/ $5 \times 10^{5}$ M$_{\odot}$)=0.29 which again implies that the Arches is likely to be a gravitationally bound cluster according to the results of Geyer \& Burkert (2001).   

\section{Conclusions}
The origin of the shallow mass functions and mass segregation observed in young starburst stellar clusters is explained by means of a model based on the efficient coalescence of pre-stellar cores and their subsequent collapse to form stars. The coalescence of cores causes a strong flattening of the turbulence generated, primordial core mass function. Once the reservoir of intermediate mass cores is depleted, and the radii of cores are significantly reduced by gravitational contraction, coalescence proceeds at a much slower pace until all cores in the whole mass range collapse to form stars. The entire process is fast, lasting for a fraction ($\sim 0.25$) of the proto-cluster cloud free-fall time, i.e., a few $10^{4}$ years. When applied to the Arches cluster, the model is able to reproduce the observed mass slopes of $\alpha \sim 2$ and $\sim 1.7$ at the intermediate and high mass ranges, respectively, as well as the peculiar bump at $\sim 6$ M$_{\odot}$. Using this model, we estimate a total mass of Arches of $1.45 \times 10^{5}$ M$_{\odot}$, which is slightly higher than the often quoted, but completeness affected, observational estimates of a few $10^{4}$ M$_{\odot}$. The star formation efficiency in this model is $\sim 30$ percent which implies that the Arches cluster is gravitationally bound according to the results of Geyer \& Burkert (2001).  

\acknowledgements 
S. D. would like to thank the conference organizers for the partial financial support which enabled him to attend the conference.

\end{document}